\documentclass[ aip, jmp, amsmath,amssymb, reprint, ]{revtex4-1}
\usepackage{amssymb}
\usepackage{amsfonts}
\usepackage{amsmath}
\usepackage{amsthm}
\usepackage{epsfig}
\usepackage{color} 
\usepackage{subfigure}
\usepackage{enumitem}   
\usepackage{hyperref}
\usepackage{cleveref}

\newcommand{\eps}{\varepsilon}

\begin{document}

\title{Collective excitability in highly diluted networks of rotators}

\author{Gabriele Paolini}
\affiliation{CY Cergy Paris Universit\'e, CNRS, Laboratoire de Physique Th\'eorique et Mod\'elisation, UMR 8089,
95302 Cergy-Pontoise , France} 
\author{Marzena Ciszak}
        \email{marzena.ciszak@ino.cnr.it}
        \affiliation{CNR - Consiglio Nazionale delle Ricerche - Istituto Nazionale di Ottica, Via Sansone 1, 50019 Sesto Fiorentino, Italy}
        \author{Francesco Marino}
        \email{francesco.marino@ino.cnr.it}
        \affiliation{CNR - Consiglio Nazionale delle Ricerche - Istituto Nazionale di Ottica, Via Sansone 1, 50019 Sesto Fiorentino, Italy}
        \affiliation{INFN, Sezione di Firenze, Via Sansone 1, 50019 Sesto Fiorentino, Italy}
\author{Simona Olmi}
\email{simona.olmi@cnr.it}
\affiliation{CNR - Consiglio Nazionale delle Ricerche - Istituto dei Sistemi Complessi, via Madonna del Piano 10, 50019 Sesto Fiorentino, Italy}
\affiliation{INFN, Sezione di Firenze, Via Sansone 1, 50019 Sesto Fiorentino, Italy}
\author{Alessandro Torcini}
\email{alessandro.torcini@cyu.fr}
\affiliation{CY Cergy Paris Universit\'e, CNRS, Laboratoire de Physique Th\'eorique et Mod\'elisation, UMR 8089,
95302 Cergy-Pontoise , France } 
\affiliation{CNR - Consiglio Nazionale delle Ricerche - Istituto dei Sistemi Complessi, via Madonna del Piano 10, 50019 Sesto Fiorentino, Italy}
\affiliation{INFN, Sezione di Firenze, Via Sansone 1, 50019 Sesto Fiorentino, Italy}

\date{\today}

\begin{abstract}
We report on collective excitable events in a highly-diluted random
network of non-excitable nodes. Excitability arises thanks to a
self-sustained local adaptation mechanism that drives the system on a
slow time-scale across a hysteretic phase transition involving states
with different degrees of synchronization. These phenomena have been
investigated for the Kuramoto model with bimodal distribution of the
natural frequencies and for the Kuramoto model with inertia and an
unimodal frequency distribution. We consider global and local
stimulation protocols and characterize the system response for different
level of dilution. We compare the results with those obtained in the
fully-coupled case showing that such collective phenomena are remarkably
robust against network diluteness.
\end{abstract}


\maketitle


\textbf{Collective excitable phenomena in a system of coupled elements may arise
either when the single nodes display excitable dynamics or not. The
latter case, more surprising, is possible when the entire ensemble of
nodes is subject to a global feedback depending self-consistently on the level of synchronization
of the network. In this case a global excitable response to an
external stimulus can be observed, which corresponds at the microscopic
level to a transient partial synchronization of the nodes.
Mathematically, this is related to hidden geometric structures that
organize the mean field trajectories in the phase-space. These events
has been observed in two paradigmatic classes of globally-coupled
rotators, namely, the Kuramoto model with and without inertia. In this paper we
analyze the robustness of the collective excitability in highly-dilute
random networks, by gradually decreasing the percentage of coupled
nodes. We consider global and local stimulation protocols and
characterize the response with respect to that achievable in a
corresponding fully-coupled network. Our findings demonstrate a
remarkable robustness of collective excitability, which we expect to
inspire new research in the study of emergent phenomena in networks of
interacting elements at the mesoscopic scale.}

\section{Introduction}

Excitable systems appear in many scientific fields, in particular they have been studied in the
context of neuroscience as simplified neural models \cite{kock1999}
as well as in cardiac dynamics for pulse propagation \cite{aliev1996}. 
Several low-dimensional models have been introduced to reproduce the
excitable properties of cells, all these models are characterized by few common
features. They all present a linearly stable fixed point, that once stimulated
with a sufficiently large perturbation display  a large excursion
in the phase space corresponding to the emission of a pulse of well-defined amplitude and duration. 
These low-dimensional slow-fast systems display a very rich dynamical repertoire,
ranging from regular spiking and bursting behaviours to their chaotic counterparts,
joined to extremely complex bifurcation structures \cite{hindmarsh1984,terman1992,wang1993,mosekilde2001,gonzalez2003,innocenti2007}.

Collective excitable responses and bursting activities have been previously reported in fully coupled networks
\cite{coombes,kawamura2011} as well as in spatially extended systems where they appear in the form of excitable waves \cite{meron_rev}
and as transient synchronization states \cite{mchaos09,dolcem}. 
However, in all these cases the observed collective features fully rely on the excitable nature of the single nodes of the network. 

A few studies \cite{so2011,skardal2014,ciszak2020,ciszak2021}
have recently shown that collective excitable dynamics can emerge also 
in networks of non-excitable units, such as rotators, in presence of a global linear feedback .
In particular, in \cite{ciszak2020} it has been investigated the effects of a global linear feedback on the dynamics of
the Kuramoto model with and without inertia. Thanks to the feedback, the system originally characterized by 
hysteretic first-order transitions  \cite{pazo2009, martens2009, tanaka1997, olmi2014, olmi2016},
reveals collective dynamical features typical of excitable models, despite a nonexcitable single-node dynamics. The origin
of these behaviors is related to the competition of the fast synchronization/desynchronization phenomena triggered by
the slow adaptation. In \cite{skardal2014}, it  has been also taken into account the case where interactions between oscillators are not mediated by a global mean field term, but occur instead through an underlying coupling network. In particular, a detailed analysis of the parameter space in a random network with power-law distributed degrees has revealed the emergence of regimes similar to those observed in the globally coupled case (bistable, synchronized, excitable, oscillatory, and incoherent states).
However, the considered network was not particularly diluted, since 
each oscillator was connected on average to the 18\% of the whole population and at least to the 10\% of oscillators.

In the present paper, starting from  \cite{ciszak2020}, we investigate how a highly diluted random network of nonexcitable nodes
can become collectively excitable thanks to a self-sustained local adaptation mechanism. In particular,
we have studied in details the excitability response of the network to perturbations by considering 
global and local stimulation protocols for different level of dilution. This analysis has
been performed for the Kuramoto model with a bimodal distribution of the natural frequencies 
and for the Kuramoto model with inertia with an unimodal frequency distribution. 
The manuscript is structured as follows. In Sec. II are described the investigated models, the indicators employed to characterize the excitability features of the system as well as the stimulation protocols.
The results of the numerical investigations are reported in Sec. III,
for the bimodal Kuramoto model (BKM) and the Kuramoto model with inertia (KMI), 
by considering the response of the network to global stimulations and to stimulation affecting only a fraction of the oscillators.
Finally in Sec. IV a brief discussion on the results is reported.

\section{Model and Indicators}\label{sec:model}

\subsection{Network Model}

We consider a heterogeneous network of $N$ rotators characterized by their
phases $\{\theta_n(t)\}$ and angular velocities  $\{\dot{\theta}_n(t)\}$, where each rotator is randomly coupled to $M$ neighbours 
via an adaptive coupling dependent on the level of synchronization among the neighbours themselves.
The evolution equations for the phases are given by
\begin{eqnarray}
\label{EQ:1a}
    m\ddot{\theta}_n + \dot{\theta}_n(t) &=& \omega_n + \frac{S_n(t)}{M} \sum_{j=1}^N 
\mathcal{C}_{nj} \sin{(\theta_{j} - \theta_n)} \\
\label{EQ:1b}
    \dot{S_n}(t) &=& \varepsilon [-S_n + K - \alpha Q_n(t)] \quad ;
\end{eqnarray}  
where $\omega_n$ is the natural frequency of the $n$-th rotator, $m$ its mass and
$S_n(t)$ an adaptive coupling controlled via a linear feedback by the modulus $Q_n(t)$
of the local Kuramoto order parameter. This is defined as
\begin{equation}
    Q_n(t) =  \frac{1}{M} \Big| \sum_{j=1}^N \mathcal{C}_{nj} e^{i \theta_j(t)} \Big| \quad ;
    \label{Rn}
\end{equation}
this quantity characterizes the level of synchronization among the $M$ neighbours of the $n$-th rotator  
and its value ranges from $Q_n=1$, if they are fully synchronized, to $Q_n \simeq \mathcal{O} (1/\sqrt{M})$ 
in case they are completely asynchronous. The gain of the feedback loop is controlled by $\alpha$ and its bandwidth 
by $\varepsilon$, usually set to $\varepsilon=0.01$.
The stationary values of the adaptive coupling ranges from $S_{FS} = K -\alpha$, for a fully synchronized case, to
$S_{AS}   \simeq K - d \frac{\alpha}{\sqrt{M}}$ for an asynchronous state, where $d$ is some constant ${\cal O}(1)$.
Therefore when the system is fully synchronized the adaptive coupling will relax towards $S_{FS}$ a coupling value
corresponding to asynchronous dynamics in the original system without adaptation, while when it is desynchronized
it will relax towards $S_{AS}$ that for $M >>1$ is definitely larger than $S_{FS}$ and corresponding to
a bistable regime in the original setup \cite{ciszak2020}. However, for small degree $M$ the latter is
no more verified, in particular it is evident that $S_{AS}$ will change sign at $M_c \simeq \left(\frac{\alpha}{K}\right)^2$ indicating
a passage from a repulsive ($ M < M_c$) to an attractive ($M > M_c$) adaptive coupling.
The matrix $\{\mathcal{C}_{nj}\}$ is an adjacency matrix, 
where $\mathcal{C}_{nj}=1$ ($\mathcal{C}_{nj}=0$)
if an undirected link exists (non exists) among rotators $n$ and $j$. The degree of each 
rotator $n$ is fixed to $M = \sum_{j=1}^N \mathcal{C}_{nj}  = \sum_{j=1}^N \mathcal{C}_{jn}$.

Since collective excitable behaviours emerge in (\ref{EQ:1a} , \ref{EQ:1b}) 
for a fully coupled network (where $M \equiv N$) whenever 
one has the coexistence of two stable regimes characterized by different levels of
synchronization \cite{skardal2014,ciszak2020},  we will focus on systems
exhibiting hysteretic synchronization transitions.
In particular, we will consider the KMI
\eqref{EQ:1a} with natural frequencies randomly selected from a Gaussian distribution (GD)
and to the usual Kuramoto model ($m=0$ in \eqref{EQ:1a}) with
a bimodal distribution of the natural frequencies (BKM). 
In order to compare with the results reported in \cite{ciszak2020}
for the fully coupled case, we will consider for the KMI $m=2$ and Gaussian distributed (GD) frequencies with zero mean and
unitary standard deviation, for the BKM bimodal Lorentzian distributed (LD) frequencies.
The LD frequencies are fixed deterministically as follows
\begin{equation}\label{EQ:2}
\resizebox{\columnwidth}{!}{$
\begin{aligned}
    \omega_j & = - \omega_0 + \Delta \tan \big[\frac{\pi}{2} \xi_j] & \xi_j & = \frac{2j - \frac{N}{2} - 1}{\frac{N}{2} + 1} & j & = 1,\dots,\frac{N}{2} \\
    \omega_j & = \omega_0 + \Delta \tan \big[\frac{\pi}{2} \xi_j] & \xi_j & = \frac{2j - \frac{3N}{2} - 1}{\frac{N}{2} + 1} & j & = \frac{N}{2}+1,\dots,N  \quad ;\\
\end{aligned}$
}
\end{equation}
where $\pm w_0$ are the position of the two peaks, each characterized by the same Half-Width Half-Maximum (HWHM) $\Delta$.
These parameters have been fixed to $w_0=1.8$ and $\Delta=1.4$.

All the numerical simulations have been performed by employing a 4-th order Runge-Kutta integration scheme with a 
time step $\delta t = 0.1$; the system has been usually investigated for a time duration $t_D = 800$.

\begin{figure*}
\includegraphics[width=0.45\linewidth]{trace_km.eps}
\includegraphics[width=0.45\linewidth]{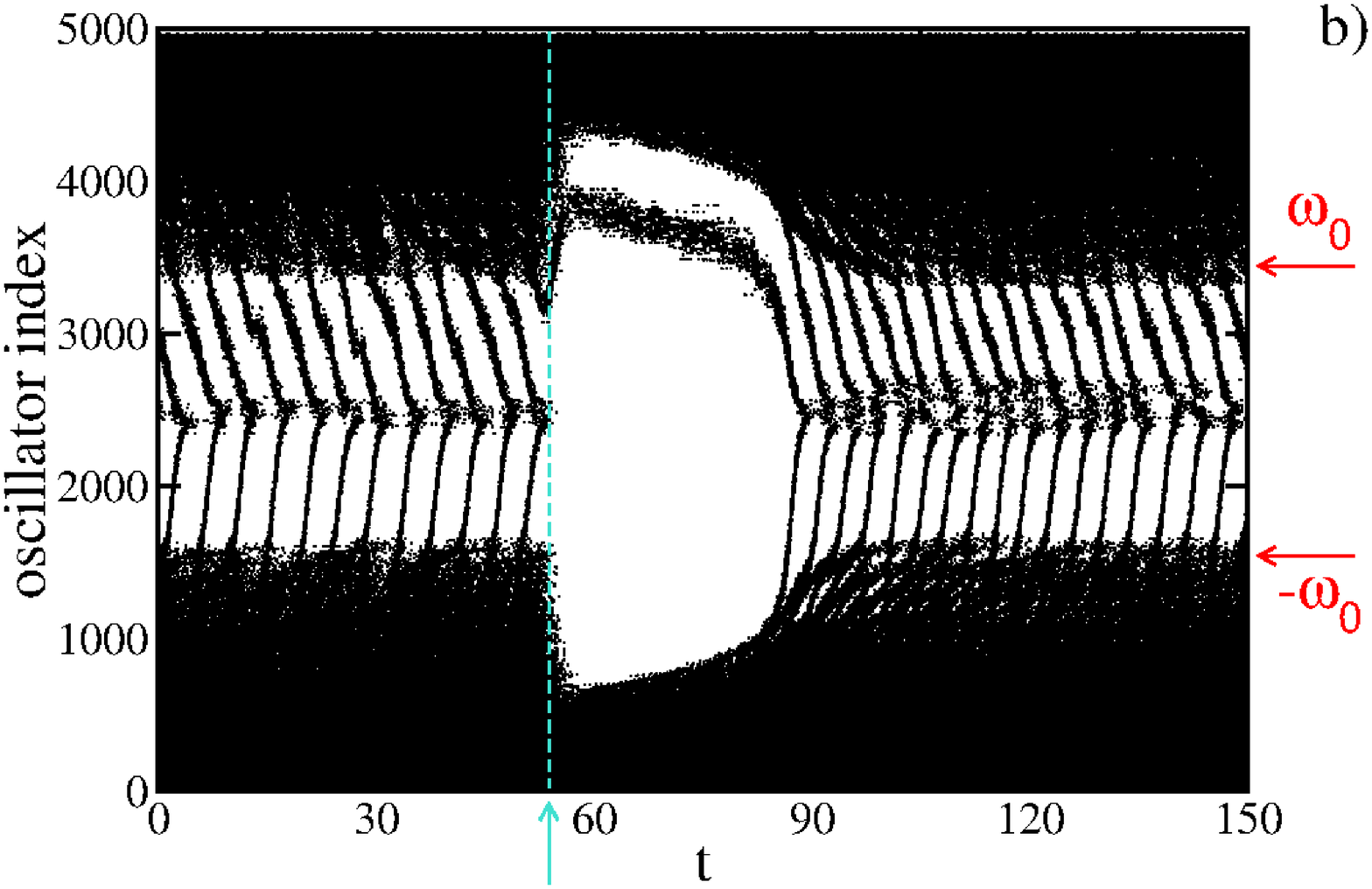}
\includegraphics[width=0.45\linewidth]{trace_kmi.eps}
\includegraphics[width=0.45\linewidth]{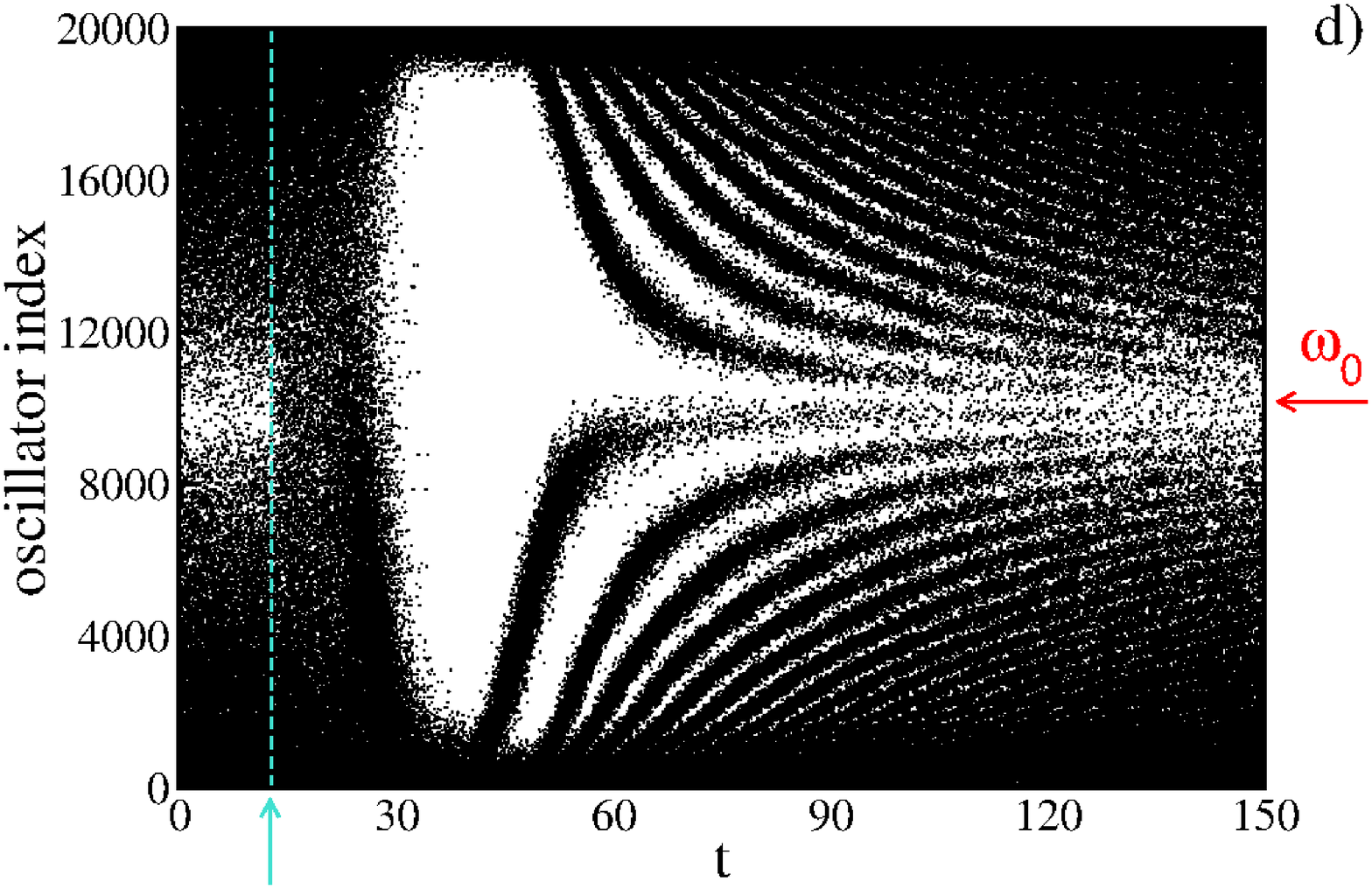}
\caption{
{\bf BKM.} a) Global order parameter $R$ vs time for a network with random connectivity (degree $M = 1000$)
in the spiking regime (black solid line). The system responses to global perturbations for different amplitudes $A$ are also displayed: a subthreshold response for $A = 0.150$ (blue solid line) and two excitable responses for $A = 0.200$ (green dashed line) and for $A = 0.375$ (red dotted line). The system response of the fully coupled network for $A = 0.375$ is shown in grey.
b) Raster plot for the diluted network obtained as a response to a perturbation of amplitude $A = 0.375$, here the oscillators are sorted by frequency in ascending order. Other parameters: $\eps = 0.01, \Delta = 1.4, \omega_0 = 1.8, \alpha = 5, K = 6.9$, and network size $N = 5000$. {\bf KMI.} c) Global order parameter $R$ vs time for a network with random connectivity ($M=50$) in the asynchronous regime (black solid line). In the same panel the system responses to global perturbations of different amplitudes $A$ are also displayed: a subthreshold response for $A = 4.2$ (blue solid line) and two excitable responses for $A = 5$ (green dashed line) and for $A = 6$ (red dotted trace). In the inset the response to a perturbation of amplitude $A=6$ is displayed for different values of the degrees: $M=50$ (red dotted line); $M=100$ (black dashed line); $M=500$ (green dots); $M=N$ (grey shaded curve).
d) Raster plot for the diluted network with $M=50$ and for a stimulation of amplitude $A = 6$, also in this case the oscillators are sorted by frequency in ascending order. Other parameters: $\eps = 0.01, \alpha = 30, K = 4.5, m=2$, $\omega_0=0$ 
and network size $N = 20000$.
The cyan arrows and the cyan dashed lines denote the time of the instantaneous stimulation.
}
\label{FIG:trace_kmi}
\end{figure*}

\subsection{Indicators}

The microscopic evolution of the oscillators have been visualized via
a sort of {\it raster plot} commonly used in the context of
neural dynamics. To be more specific, for each oscillator we
depict a dot in correspondence of the time instant the oscillator's phase
crosses a fixed threshold, in our case the threshold has been fixed
to $\theta_{th} = 0 $.

In order to characterize the collective dynamics of the system we will focus on 
the global Kuramoto order parameter  \cite{acebron2005}
\begin{equation}
R(t) =  \frac{1}{N} \Big| \sum_{j=1}^N e^{i \theta_{j}(t)} \Big| \quad .
\end{equation}
In absence of coupling adaptation, one typically observes asynchronous (partially synchronous) regime characterized by $R \simeq \mathcal{O} (1/\sqrt{N})$ 
(finite $R$) for small (large) coupling strenghts \cite{kuramoto}.

\subsection{Stimulation Protocols}

In this work, since we want to analyze the excitable properties of the system, 
we are interested in observing the response of the system to perturbations. In particular,
we will perturb instantaneously the adaptive coupling term $S_n(t)$  of the oscillator $n$
by increasing its value of a constant amount $A$ at time $t$: i.e. $S_n(t^{+}) = S_n(t^{-}) + A$.
If all oscillators are perturbed at the
same time this corresponds to perform a \textit{global stimulation} of the system,
instead if only a percentage $P$ of oscillators is perturbed this will be termed
\textit{local stimulation}. 

For local stimulations we usually stimulate the oscillators irrespectively of their natural frequencies for the KMI,
while for the BKM by starting by the central node with index $N/2$ and then 
moving symmetrically towards nodes with larger and smaller indeces (protocol PT0).
For the BKM besides this protocol, we consider also two other different stimulation protocols
defined as follows: (PT1) the oscillators are stimulated by considering first the ones
with natural frequencies $\omega_i \simeq |w_0|$ 
in proximity of the two peaks of the frequency distribution function
and then symmetrically the oscillators with
larger values of $|\omega_i - \omega_0|$ and $|\omega_i + \omega_0|$;
(PT2) is the same protocol as (PT1) but referred only to one peak,
namely the one located at $-\omega_0$, therefore the oscillators
are stimulated asymmetrically starting from the ones in proximity of only
one of the two peaks of the distribution. 

The excitable response of the system to the stimulations have been characterized in terms of the global Kuramoto order parameter $R(t)$, a population burst is identified whenever $R(t)$ overcomes a threshold value $R_{th}$ within
a given time window $W_{test}$ after the perturbation deliverance. Furthermore, we measured the maximum 
value $R_m$ at the burst peak as well as the time $T_m$ needed to reach such peak after the stimulation.
For the BKM we fixed $R_{th}=0.4$, due to the presence of spiking activity, and $R_{th}=0.2$ for the KMI,
in both cases $W_{test}=200$.

\section{Results}

We will consider parameter values for which the fully coupled adaptive networks ($M=N$) display collective excitable properties for BKM and KMI, as reported in \cite{ciszak2020}, and we will examine the response of the randomly diluted systems ($M << N$) to global and local stimulations by varying $M$ as well as the percentage $P$ of stimulate oscillators.

In particular, for the BKM, by following \cite{ciszak2020} we analyze a situation where the fully coupled system displays collective oscillations, characterized by the
alternation of partially synchronized phases with abrupt desynchronization events, usually termed {\it spikes} \cite{wang1993}. 
This behaviour persists also for diluted systems 
as shown in Fig. \ref{FIG:trace_kmi} (a) (black trace) for $K=6.9$, $\alpha=5$ $M=1000$ and $N=5000$ and this is the regime that we will
analyze in the following for the BKM. For what concerns the collective excitable properties of the KMI, we will consider a purely asynchronous phase, where $R(t)$ show irregular fluctuations of amplitude $\mathcal{O} (1/\sqrt{N})$, this occurs for $\alpha=30$ and $K=4.5$ as shown in Fig. \ref{FIG:trace_kmi} (c) (black solid line) for $M=50$ and $N=20000$.

\subsection{Global Stimulation}\label{sec:perturb_all}

\subsubsection{Bimodal Kuramoto Model}

Let us first consider the BKM in the spiking regime, where
the corresponding time evolution of $R(t)$ is reported as a solid line in Fig. \ref{FIG:trace_kmi} (a). 
In this regime the system displays collective excitability when globally stimulated despite the oscillators are randomly connected
with high dilution $M << N$. Indeed, small contemporary perturbations of all the feedback variables $S_n(t )$ elicit rapidly decaying 
responses in $R(t)$ (blue trace in Fig. \ref{FIG:trace_kmi} (a)), while sufficiently large stimuli induce 
a synchronized activity in a large part of the network (a so-called {\it burst}) characterized by an evolution of $R(t)$ corresponding to a large amplitude oscillation with a well-defined shape, amplitude, and duration (green trace).
Furthermore, a larger perturbation gives rise to a bigger burst (red dotted line)
that involves more synchronized oscillators and approaches an asymptotic shape essentially corresponding to that
observed in the fully coupled network for the same perturbation amplitude $A_0$ (grey curve). This despite the
system being indeed diluted with a clustering coefficient $c = M/N = 20 \%$ \cite{newman2003}.

The mechanism leading to the emergence of a burst can be better understood by examining the corresponding raster plot
reported in Fig. \ref{FIG:trace_kmi} (b). In the period before the moment of the stimulation (denoted by 
a cyan arrow and a dashed cyan line) one observes a spiking regime characterized by standing waves in the raster plot propagating
from oscillators with natural frequencies $\pm \omega_0$ towards the oscillators with zero frequency.
Immediately after the stimulation a wider group of oscillators with natural frequencies in the interval 
$[-\omega_0 - \Delta;\omega_0 + \Delta]$ phase lock, while the remaining oscillators with larger
natural frequencies in absolute value are not entrained to the big synchronized cluster. This is the origin
of the burst as measured by the order parameter $R(t)$. The number of the synchronized oscillators slowly
decreases in time, but it remains essentially unmodified for a duration $t_d \simeq 30$, soon after the burst
disappears and the standing waves re-emerge in the network.

It is important to remark that spikes are hardly discernible for $M <50$ from 
the fluctuations of $R$ due to the network sparsity, and this sets for the BKM the minimal
degree that can be analyzed to $M=50$ corresponding to a clustering coefficient $c = M/N = 1 \%$.

\begin{figure*}
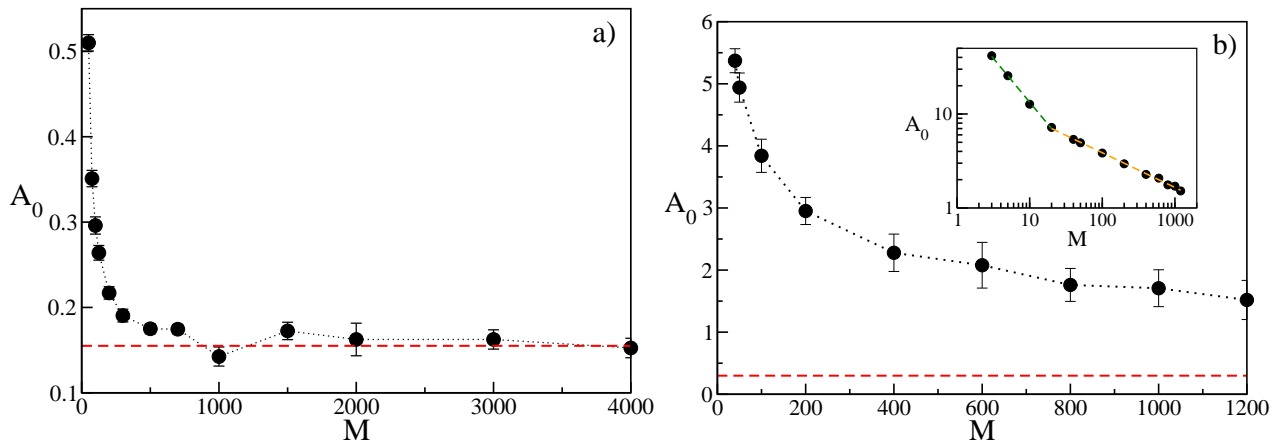

\includegraphics[width=0.48\linewidth]{soglia_minima_eccitabilita_km.eps}
\includegraphics[width=0.45\linewidth]{soglia_minima_eccitabilita_kmi.eps}
\caption{
Minimal perturbation amplitude $A_{0}$ required to observe a burst as a function of the degree $M$ :
a) {\bf BKM} and b) {\bf KMI}.
The red dashed lines represent the values of the minimal perturbation amplitude 
for the corresponding FC networks. In the inset of panel (b) the green (orange) dashed lines
refer to a power law fit $a/M^\beta$ to the data in the interval $3 \le M \le 20$ ($20 \le M \le 1200$)
with an exponent $\beta \simeq 0.93$  ($\beta \simeq 0.37$).
For the BKM for $M < 1000$, we averaged $A_{0}$ over $50-100$ network realisations, while for greater values of $M$ we averaged over $10$ realisations due to higher computational time, for the KMI we always averaged over 20 network realizations. In both panels the error bars correspond to the standard deviation of the mean. Parameters as in Fig.\ref{FIG:trace_kmi}, apart $K=4.0$ for the KMI.}
\label{FIG:soglia_minima}
\end{figure*}

\subsubsection{Kuramoto Model with Inertia}

In the case of the KMI the unperturbed state is asynchronous at variance with the BKM, as evident
from the evolution of $R(t)$ reported as a solid black line in
Fig. \ref{FIG:trace_kmi} (c). However, also in this case the systems displays collective excitability 
when globally stimulated even in the very diluted case, namely we considered $M=50$ corresponding to a clustering coefficient
$c=0.25 \%$ for $N=20000$. Indeed,
small perturbations of amplitude $ A =  4.2$ elicit no collective response (blue solid line), while
larger perturbations ignite bursts, as shown in Fig. \ref{FIG:trace_kmi} (c) for $A=5$ (green dashed line)
and $A=6$ (red dotted line). This demonstrates the existence of a threshold value $A_0$ which should be overcome
to observe an excitable response when globally perturbing the system. Furthermore, 
as shown in the inset of Fig. \ref{FIG:trace_kmi} (c) the shape of the burst
approaches that obtained in the globally coupled case by increasing $M$ and already for $M=500$ (green dots)
the asymptotic profile corresponding to $M=N$ is essentially recovered.

In this case the collective response of the system to a stimulation is quite different with respect to the BKM
as shown in Fig.\ref{FIG:trace_kmi} (d). In particular, there is a transient period of duration $t_T \simeq 15$
before the onset of the burst that was definitely shorter in the BKM case. Once more during the burst a large part of the rotators are phase locked, at variance with the BKM in this case almost all the rotators are involved in the burst. At its disappearence we observe that due to inertia the rotators with natural frequencies in proximity of the peak of the distribution located at $\omega_0=0$ 
relax faster to the asynchronous regime than those far away in the distribution itself characterized by
high natural frequencies $|\omega_i|$.

\subsubsection{Comparison of BKM and KMI analysis}

As a first analysis we investigate the value of the minimal perturbation amplitude $A_0$ needed to ignite a population burst
as a function of the network degree $M$. The results of the analysis are reported in Fig.\ref{FIG:soglia_minima} for the BKM and KMI.
While in the BKM we cannot consider $M < 50$ otherwise the spiking activity
will be no more discernible from the background noise, for the KMI even for vanishingly small $M$ the dynamics remains
asynchronous. In this latter case we observed that we have a collective response even for $M=3$ with a sufficiently large
perturbation $A_0$ (as shown in the inset of Fig. \ref{FIG:soglia_minima} (b)), however no response was observable for 
$M=1$ and 2 even for $A_0$ extremely large. This is probably due to the existence or not of a giant component
in our random network. As a matter of fact for the random network we are
analysing, where the degree is always exactly $M$, we have the emergence of a giant component for the network
via a phase transition occuring at $M=2$, while for an Erd\"os-Renyi network the giant component appears at $M \ge 1$ \cite{molloy1995,newman2003}.

For the BKM the minimal amplitude $A_0$ approaches the FC value with a power law decay
characterized by an exponent $\beta \simeq 1.30$, while for the KMI the approach to the FC case seems definitely
more complex. In this case, $A_0$ decreases quite rapidly at small $M \le 20$ decaying as $\propto 1/M$ while at larger $M$ we still observe a power-law decay of $A_0 \propto M^{-\beta}$ but with a definetely smaller exponent $\beta \simeq 0.4$. This crossover from a fast to a slower decay of $A_0$ with the degree is clearly visible in
the inset of Fig. \ref{FIG:soglia_minima} (b). The crossover is probably related to the fact that $S_{AS}$ changes sign,
from positive to negative, for $M \simeq 40$, not far from the expected value $M_c = \left(\frac{\alpha}{K}\right)^2 = 56.25$.
At $M \simeq 40$ we pass from a situation where in the stationary case the effective coupling among the rotators is repulsive 
to the usual situation where the coupling is attractive. Indeed at $M \leqslant 40$ definitely higher perturbation amplitudes
$A_0$ are required to recruit large part of the oscillators in a collective burst.  Therefore it is reasonable to expect that $A_0$ decreases
faster with increasing $M$ for $M \le M_c$ than at larger values.

The fact that the exponent $\beta$ is definitely larger for the BKM than for the KMI when approaching the FC case
can be probably related to the fact that it is easier to elicit a population burst in a system presenting already 
a partial synchronization associated to the standing waves (spiking activity), than in a system that reveals an asynchronous dynamics.

\begin{figure*}
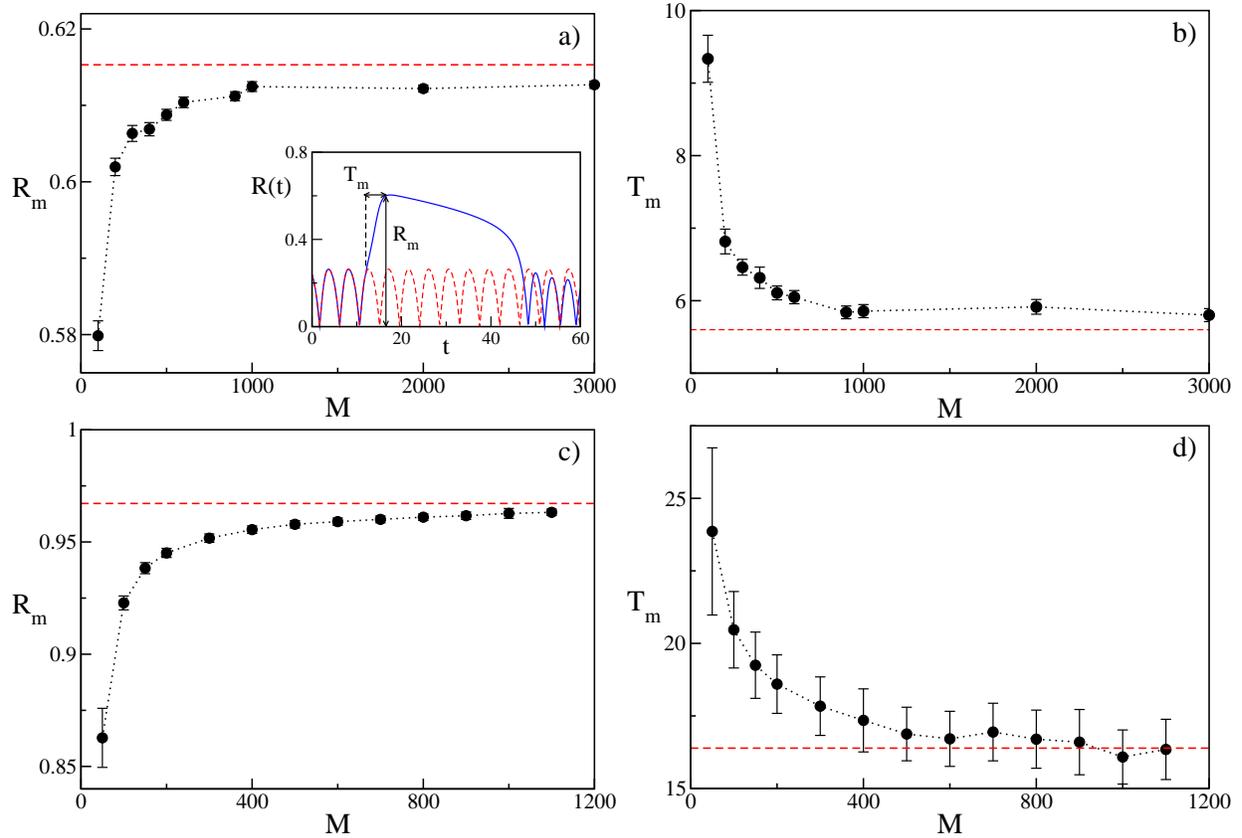

\includegraphics[width=0.45\linewidth]{Rmax_km.eps}
\includegraphics[width=0.45\linewidth]{tmax_km.eps}
\includegraphics[width=0.45\linewidth]{Rmax_kmi.eps}
\includegraphics[width=0.45\linewidth]{tmax_kmi.eps}
\caption{{\bf BKM.} System response to a global stimulation of amplitude $A = 0.4$: a) maximum value of the order parameter $R_m$ 
and b) time $T_m$ needed to reach it as a function of the degree $M$.
{\bf KMI.} System response to a global stimulation of amplitude $A = 6$: 
 c) maximum value of the order parameter $R_m$ 
and d) time $T_m$ needed to reach it as a function of the degree $M$.
The red dashed lines represents, in all panels, the values obtained for the FC network  averaged over 10 
different initial conditions. Data are averaged over $100$ realisations and the error bars correspond to the 
standard deviation of the mean. Other parameters are fixed as in Fig.\ref{FIG:trace_kmi},apart $K=4.0$ for the KMI.}
\label{FIG:variM_kmi}
\end{figure*}

Furthermore, we characterize the emergence of the burst in terms of two indicators associated
to the global order parameter $R(t)$ shown in the inset of Fig. \ref{FIG:variM_kmi} (a): the value of the maximum $R_m$ reached by the order parameter once stimulated and the time $T_m$ needed to reach such maximum after the stimulation. 
As a first analysis we consider diluted systems with different degrees $M$ and we globally stimulate the system with perturbations of amplitude $A$ sufficiently large to lead to a population burst for the network with the smallest considered $M$. In particular,
for the BKM (KMI) the minimal considered degree is $M=100$ ($M=50$) and $A=0.4$ ($A=6.0$).
The each fixed $M$ we measured the average value of $R_m$ and $T_m$ obtained by considering
100 different realizations of the random network. The results are reported in Fig. \ref{FIG:variM_kmi} (a,b) for the BKM and 
Fig. \ref{FIG:variM_kmi} (c,d) for the KMI.

In both cases we observe that for increasing $M$ the value of $R_m$ approaches those obtained in the fully coupled
case $R_m^{(FC)}$ for the same stimulation amplitudes $A$ with a scaling law consistent with $R_m^{(FC)} - a/M^\beta$ with
$\beta \simeq 1.0 - 1.1$. Analogously by increasing $M$ the time needed to reach the maximum of the burst decreases 
towards the FC value $T_m^{(FC)}$ with a power law decay consistent with $T_m^{(FC)} +b/M^\beta$ where $\beta \simeq 1.0 - 1.1$.
These results suggest that the finite size corrections due to the dilution for the measured quantities should vanish as $\propto M^{-1}$
by approaching the FC limit.

\begin{figure*}
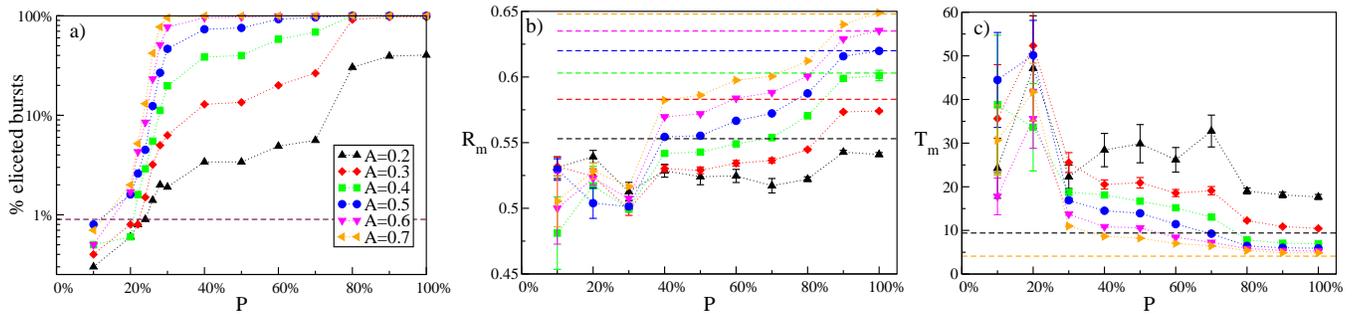

\includegraphics[width=0.33\linewidth]{nburst_amp_e_perc_variabili_M200_RS.eps}
\includegraphics[width=0.33\linewidth]{Rmax_amp_e_perc_variabili_M200_RS.eps}
\includegraphics[width=0.32\linewidth]{tmax_amp_e_perc_variabili_M200_RS.eps}
\caption{{\bf BKM} System response to stimulations of various amplitudes applied to a different percentage $P$ of oscillators, for a fixed degree ($M = 200$). a) Percentage of elicited bursts vs the percentage $P$ of perturbed oscillators.
Different curves represent simulations for different perturbation amplitudes $A$; the dashed horizontal line represents the percentage of spontaneous emissions. 
b) Maximum value of the order parameter $R_m$, measured at the burst peak, vs $P$, for different $A$ values.
c) Times values $T_m$ the system takes to reach the burst peak vs $P$ for different $A$ values.
The dashed lines in panels b), c) represent the values obtained from the simulations of the FC networks for the corresponding stimulation amplitude $A$. The colors identify the different $A$ values as reported in the legend in panel (a).
In all panels data are averaged over $1000$ different realisations and the error bars correspond to the standard deviation of the mean. See Fig.\ref{FIG:trace_kmi} for the parameters.}
\label{FIG:M200}
\end{figure*}

\section{Local Stimulation}\label{sec:perturb_partial}

In this section we extend the analysis previously performed to take into account the response of the system to local stimulations. In particular we investigate the response of the systen
to perturbations involving different percentages $P$ of oscillators for different level of network dilution
(measured by the degree $M$) and for different amplitude stimulations $A$.

\begin{figure}
\includegraphics[width=0.90\linewidth]{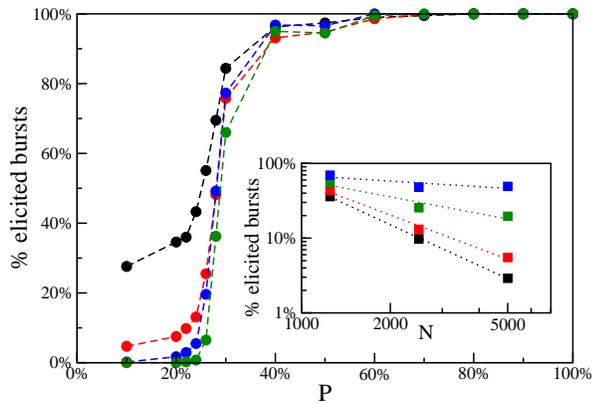}
\caption{{\bf BKM} Finite size effects  for fixed degree $M=200$ and amplitude of perturbation $A=0.5$.
Percentage of elicited bursts vs the percentage $P$ of perturbed oscillators for various system sizes:
$N=1250$ (black circles), $N=2500$ (red ones), $N=5000$ (blue ones) and $N=10000$ (green ones).
The dashed lines are a guide for the eyes.
In the inset are reported the percentage of elicited bursts for different system sizes
for various percentage $P$ of stimulated oscillators : namely, $P=22 \%$ (black squares),  
$P=24 \%$ (red ones) , $P=26 \%$ (green ones) and $P=28 \%$ (blue ones).
The dotted lines are power-law fitting to the data of the type $a +b N^{-\gamma}$,
the values of the estimated exponents are : $\gamma \simeq 1.82$ ($P=22 \%$),
$\gamma \simeq 1.49$ ($P=24 \%$), $\gamma \simeq 0.75$ ($P=26 \%$) and $\gamma \simeq 0.25$ ($P=28 \%$).
The data are averaged over $1000$ different realisations, apart for $N=10000$, where
only 200 realizations are considered. Other parameters as in Fig.\ref{FIG:trace_kmi}.}
\label{FIG:Finite}
\end{figure}

\subsubsection{Bimodal Kuramoto Model}

Let us first consider the BKM, in this case we fix the degree to $M=200$ and we measure for different percentage $P$
of stimulated oscillators and a certain stimulation amplitude $A$ the percentage of bursts elicited by the stimulation
by considering 1000 different realizations of the network.  The corresponding results are reported in Fig. \ref{FIG:M200} (a).
As a first analysis we estimated the percentage of bursts emitted in absence of any stimulation by varying $P$, the bursts spontaneously emitted correspond to the $0.9 \%$ of the considered realizations
(dashed horizontal line in Fig. \ref{FIG:M200} (a)). Therefore below this percentage the system reponse
will be considered not reliable. As shown in Fig. \ref{FIG:M200} (a) for any considered value of $0.2 \le A \le 0.7 $ 
the number of elicited bursts is below this threshold whenever the number of stimulated oscillators is below the $10\%$ of
the nodes of the network, corresponding in this specific case to $500$  oscillators.

From Fig. \ref{FIG:M200} (a) we observe that the number of elicited bursts increases with
the percentage $P$ of stimulated oscillators for any considered amplitude $A$, however
for $A=0.2$ even by stimulating all the oscillators we obtain bursts only in the
$40 \% $ of the realizations. Indeed, this is related to the fact that $A=0.2$
is slightly below the minimal amplitude of stimulation needed to observe a burst for $M=200$,
that is $A_0 = 0.217 \pm 0.007$.

For larger amplitudes $A > 0.2$ it is possible to observe a burst
for each delivered stimulation (corresponding to 100 $\%$ of elicited bursts),
whenever  $P$ is larger than a critical value $P_f= P_f(A)$ which decreases quadratically with $A$ 
($P_f(A) \propto A^{-2}$) for $A \geq 0.5$.
 
In Fig. \ref{FIG:M200} (b) and (c) we report the value of $R_m$ and $T_m$ averaged over
all observed bursts for the same amplitudes and $\%$ of stimulated oscillators analysed in panel (a).
As a general remark we observe that $R_m$ ($T_m$) grows (decreases) monotonically with the presentage of
stimulated oscillators only when this value is beyond the $30 \%$. 
As we will see in the following this is due to finite size effects.
Furthermore, for sufficiently large amplitudes ($A \ge 0.4$) the maximal
value of the burst tends towards that measured in the FC network $R_m^{(FC)}$ for the same amplitude,
as expected $R_m^{(FC)}$ is reached when all the oscillators are simultaneously stimulated.
The same is essentially true for $T_m$. For low amplitudes $A \le 0.3$ the asymptotic shape
observed in the FC network is never reached, this becasue the stimulation amplitudes are smaller 
or of the order of the minimal one required to elicit a burst in the FC network.

All the analysis for the BKM have been sofar performed for $N=5000$, it is important in at least one
case to evaluate the relevance of the finite size effects. Therefore for fixed degree $M=200$
and perturbation amplitude $A=0.5$ we examine the response of the network for different
system sizes : namely, $N=1250$, 2500, 5000 and 10000. The results of this analysis, reported
in Fig. \ref{FIG:Finite}, show clear finite size effects for $N=1250$ and $N=2500$ at
percentage $P$ of stimulated oscillators below $30 \%$. As shown in the inset the
finite size effects become less and less relevant for increasing $P$, in particular
for  small $P$-values ($P < 30\%$) the percentage of elicited bursts displays a power-law decay $\simeq N^{-\gamma}$
with an exponent $\gamma$ strongly dependent on $P$. A linear extrapolation of the decay of this exponent
$\gamma = \gamma(P)$ provide us with a critical value $P_c(A=0.5) \simeq 28.7 \%$ below which no burst
can be elicited for sufficiently large system sizes. This represents an activation threshold 
to observe excitable properties in the system in the
thermodynamic limit.  Obviously this threshold depends on value of $A$, but it is  a generic feature
of the studied random network present for any perturbation amplitude $A \ge A_0$.
On the other hand, above this threshold value  the finite size
effects can be assumed to be negligible. Furthermore, since for $P=30\%$ of stimulated oscilators
we have elicited bursts already in the 75-77 \% of cases, we can affirm that the transition from no elicited bursts
to a finite number of bursts should occur in an extremely narrow interval of $P$ values in the thermodynamic limit.
The analysis of the nature of this transition, continuous or discontinous, is left to future analysis.

\begin{figure*}
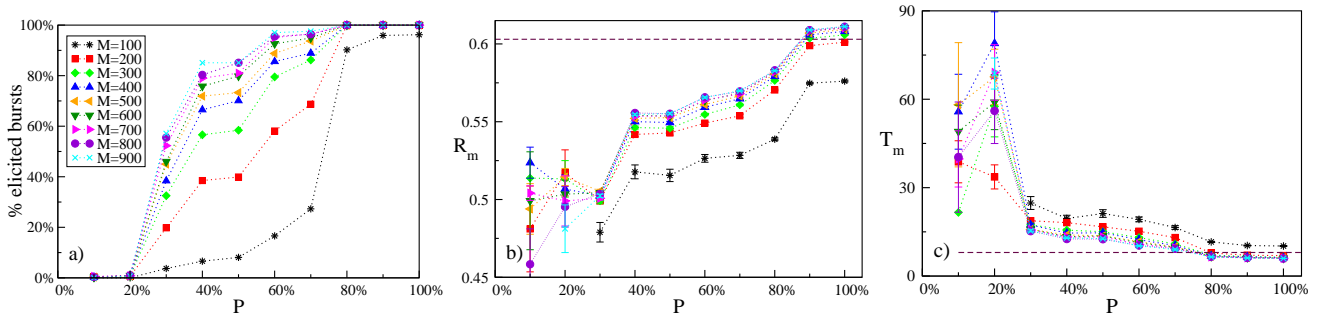

\includegraphics[width=0.325\linewidth]{nburst_km_variM.eps}
\includegraphics[width=0.31\linewidth]{Rmax_km_variM.eps}
\includegraphics[width=0.32\linewidth]{tmax_km_variM.eps}
\caption{{\bf BKM} System response to stimulations with fixed amplitude ($A = 0.4$) for different degrees $M$: percentage of elicited bursts (a), $R_m$  (b) and  $T_{m}$ (c) versus the percentage $P$ of perturbed oscillators. 
The black dashed lines represent the values for the FC network $M=N$. 
Data are averaged over $1000$ different realisations of the networks and the error bars correspond to the standard deviation of the mean. See Fig.\ref{FIG:trace_kmi} for the other parameters.
}
\label{FIG:part_variM_km}
\end{figure*}

Let us now examine the response of the system for a fixed perturbation amplitude for various degree values  $M \in [100:900]$
corresponding to clustering coefficients in the range $c \in [2\%:18\%]$ by varying the percentage $P$ of stimulated oscillators.
The corresponding results are reported in Fig. \ref{FIG:part_variM_km} (a). 
The perturbation amplitude $A = 0.4$ has been chosen in order to be larger than the minimal excitability threshold 
$A_0$ already at $M = 100$. As a first  remark we observe that below a stimulation threshold
involving at least a percentage $P_m=  20 \%$ of oscillators almost no burst is elicited. 
Furthermore, the increase of $M$ leads
to a larger number of emitted bursts, once fixed the $\%$ of stimulate oscillators $P$. However, we observe
a 100 \% of elicited bursts only when at least a percentage $P_f =80 \%$ of oscillators is perturbed irrespectively of $M$,
apart for $M=100$ where a one to one correspondence between stimulation and burst occurrence is never achieved for this 
stimulation amplitude.
Similarly for $M \ge 200$ the quantities $R_m$ and $T_m$, characterizing the burst, approach their FC values
for increasing $P$, values that are reached whenever $P \ge P_f$ (see Fig. \ref{FIG:part_variM_km} (b,c)).

\begin{figure*}
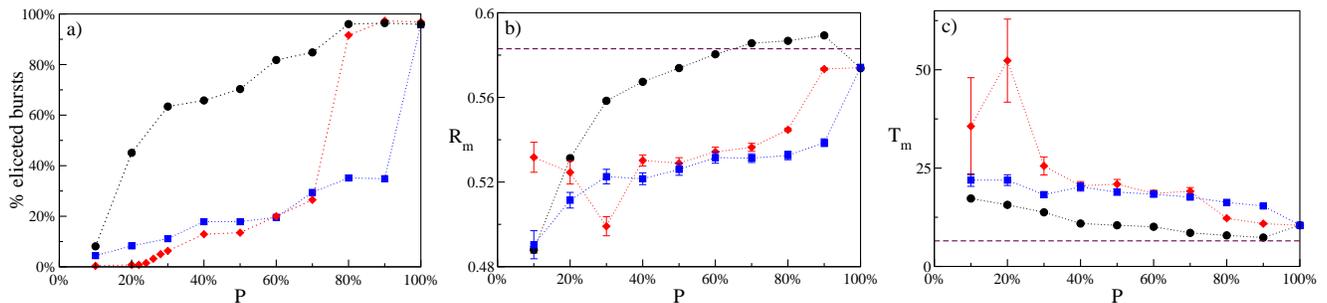

\includegraphics[width=0.32\linewidth]{confronto.eps}
\includegraphics[width=0.32\linewidth]{confronto_rmax.eps}
\includegraphics[width=0.32\linewidth]{confronto_tmax.eps}
\caption{{\bf BKM}
Comparison among different perturbation protocols: protocol PT0 (red line with diamonds);
protocol PT1 (black line with circles); and protocol PT2 (blue line with squares). Percentage of elicited bursts (a), $R_m$  (b) and  $T_{m}$ (c) versus the percentage $P$ of perturbed oscillators. The black dashed lines represent the values for the FC network $M=N$. Data are averaged over $1000$ realisations of the network and the error bars correspond to the standard deviation of the mean. $A = 0.4$, $M = 200$, other parameters as in Fig.\ref{FIG:trace_kmi}.
}
\label{FIG:comparison_protocol}
\end{figure*}

In order to understand if the order of stimulation of the oscillators is relevant, we consider 
besides the standard protocol PT0, employed in the rest of the article, also 
two other protocols PT1 and PT2 previously defined in SubSection II-C.
In particular, we analyze the excitable
response of the network to these local stimulations protocols 
for fixed degree $M=200$ and fixed amplitude stimulation, namely $A=0.4$.
The results of this analysis are displayed in Fig. \ref{FIG:comparison_protocol}.
It is clear that the most effective protocol to induce excitable bursts is $P1$, this
because by stimulating the oscillators with natural frequencies in correspondence of
the peaks of the frequency distribution favours the formation of synchronized clusters.
On the other hand the protocol PT2 is the less effective since the stimulations of 
oscillators with natural frequencies in proximity of only one peak results in an asymmetric 
recruitement of oscillators which do not favour global synchronization.
The protocol PT0 shares a similar problem by recruiting first the oscillators located 
in the minimum of the probability distribution function between the two peaks,
indeed the results are quite similar to PT2 apart at very large
percentage $P$ of stimulated oscillators (beyond 70 \%), where PT0 performs better.

This is evident also by the values obtained for $R_m$
(shown in Fig. \ref{FIG:comparison_protocol} (b)) which are definitely smaller for PT0 and PT2 protocols
with respect to PT1 for the same percentage $P$ of stimulated oscillators. Indeed the values of $R_m$ for PT1 are already 
comparable to the FC result for $P=40 \%$.
Also the activation of the oscillators  involved in the burst is faster for PT1 then
for the other two protocols , as shown in Fig. \ref{FIG:comparison_protocol} (c).

These results clearly indicate that a stimulation of oscillators guided by their natural
frequency distribution is the optimal one to elicit more bursts involving more oscillators.

\subsubsection{Kuramoto Model with Inertia}

\begin{figure*}
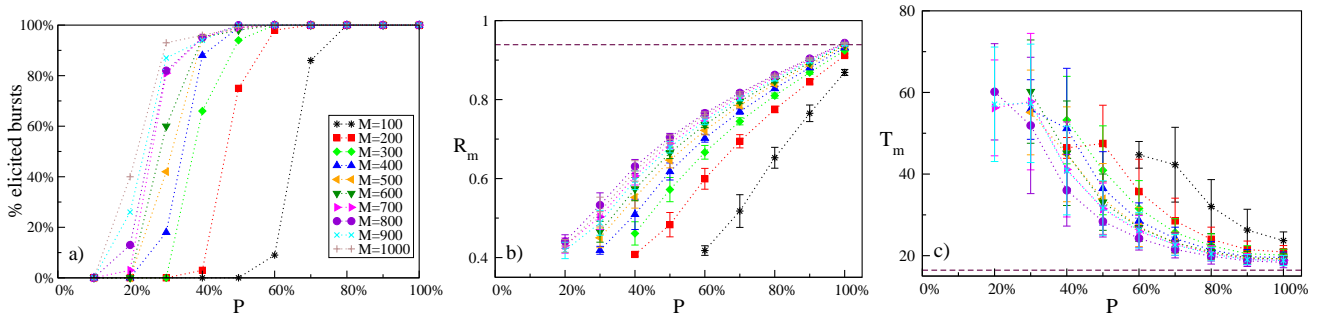

\includegraphics[width=0.325\linewidth]{nburst_kmi_variM.eps}
\includegraphics[width=0.31\linewidth]{Rmax_kmi_variM.eps}
\includegraphics[width=0.32\linewidth]{tmax_kmi_variM.eps}
\caption{{\bf KMI} System response to local stimulations of fixed amplitude ($A = 6$) for different degrees $M$:
percentage of elicited bursts (a), $R_m$  (b) and  $T_{m}$ (c) versus the percentage $P$ of perturbed oscillators. 
The black dashed lines represent the  FC value. The data refer to an average over $100$ different realizations of the network
and the error bars correspond to the standard deviation of the mean.
See Fig.\ref{FIG:trace_kmi} for the other parameters.  
}
\label{FIG:part_variM_kmi}
\end{figure*}

Finally we present the results for the local stimulation of KMI, where we investigate the response of the system for a fixed perturbation amplitude, while varying the degree $M$ of the considered networks (see Fig. \ref{FIG:part_variM_kmi}).
We have chosen an amplitude $A=6$ definitely larger than the minimal amplitude $A_0$ needed to ignite
a burst for the KMI for any $M \ge 100$. At variance with the BKM case, the minimal percentage $P_m$ of stimulated rotators to observe a burst as well as the critical percentage $P_f$ required to observe the 100 \% of elicited bursts strongly depends on $M$.
This is true at least for $M \le 500$, corresponding to clustering coefficients $ c \le 2.5 \%$, definitely
smaller than those examined in the BKM case where $ c \ge 2 \%$.
Indeed for $M=100$ one needs to stimulate the 50 \% of the population to elicit a burst, while $P_m$ decreases to 20\% 
only for $M \geq 700$. Furthermore, we observe in this case that $P_f \simeq P_m + 30\%$, therefore
after the bursts' activation a moderate increase in $P$ leads to a one to one
response to the stimulation. However, this is not joined with the achievement of the characteristcs of the burst
displayed in the FC case. As shown in Fig. \ref{FIG:part_variM_kmi} (b) and (c) the approach to the FC values
for $R_m$ and $T_m$ is quite gradual and these are achieved only when essentially the whole network is stimulated.
These effects are clearly due to the inertia present in the model that makes the single rotators more
resilient to stimulation.

\section{Conclusions}

Excitability is an important property of many living cells such as
neurons whereby a large, rapid change in the membrane potential is
generated in response to a very small stimulus.
While its mechanisms are reasonably well understood at the level of the
single elements, the origin of collective excitable phenomena in
mesoscopic populations are less clear.

Of particular interest is the case of adaptive networks of globally
coupled non-excitable units, in which the emergent macroscopic dynamics
cannot be deduced in any way from the individual properties of the
nodes. A key role here is played by the adaptive feedback, which allows
for the coupling to slowly evolve as a function of the degree of
synchrony of the system, thus the network is permanently driven across a 
hysteretic phase transition. The result are collective slow-fast dynamics,
such as bursting and excitability via canard explosions, that arise even
for small population sizes \cite{ciszak2021}.

In this work, we provided evidence of emergent collective excitability in
highly-diluted random networks of rotators. This phenomenon has been
studied for the Kuramoto model with and without inertia, considering
different global and local stimulation protocols and different levels of
dilution. 

In the case of global stimulation, we can observe excitable responses
down to dilution level of 1 \% for the BKM and 0.015 \% for the KMI
by increasing the perturbation amplitude. Quite astonishngly the KMI 
with linear feedback can exhibit excitable responses down to a degree $M > 2$
corresponding to the emergence of the giant component in the corresponding network 
without adaptation \cite{molloy1995,newman2003}.
This is in line with what reported in \cite{olmi2014} where it was
shown that the hysteretic transition for the KMI persists down to a degree $M=5$, but for a
smaller network size, namely $N=5000$, where finite size fluctuations
are larger than in the network of size $N=20000$ here considered.
As a general effect the dilution reduces the level 
of synchronization $R_m$ achieved during the population burst and increases the time $T_m$
needed to reach the peak of the burst for the same stimulation amplitude.
In the limit $ M \to N$ the fully coupled results are recovered
and the finite size corrections due to the dilution vanish as $\propto 1/M$ 
by approaching the FC limit.
 
For local stimulations, where only a certain percentage $P$ of rotators
is stimulated the collective excitability emerges only for $P \ge 20 - 30 \%$.
A detailed finite size analysis performed for the BKM at fixed dilution and
perturbation amplitude suggests 
the existence of a phase transition from a non excitable to an excitable
regime occurring for a finite percentage of stimulated oscillators $P_c$. 
The exact nature of the transition, continuous or discontinuous, is left to future analysis,
however it appears to be quite abrupt. Furthermore, for $P > P_c$ we observe that 
the excitable response of the system becomes more and more reliable
and finally for $P \ge P_f$ elicited bursts are observable for any
considered network realization. The threshold value $P_f$ depends slightly on the level of
dilution, however it decreases strongly by increasing the amplitude $A$ 
of the perturbation: namely, $P_f \propto 1/A^2$ for $A \ge 0.5$ and $M=200$.
 
Another interesting aspect revealed by the analysis of the BKM is that the
more effective way to induce collective excitable responses is to stimulate 
the oscillators accordingly to the distribution of their natural frequencies.
In the case of bimodal distributions this amounts to stimulate symmetrically the
oscillators with frequencies in correspondence of the two peaks of the probability
distribution function.

The inertia plays a fundamental role when local stimulations are considered,
indeed at variance with the BKM case for a fixed stimulation amplitude 
the minimal percentage $P_m$ of rotators to stimulate in order to observe 
at least one burst as well as the threshold value $P_f$ strongly depends on $M$.
Furthermore, an excitable response analogous to that of a FC network is
achieved only when the entire network is stimulated simultaneously, while for the BKM this
was already observable in most cases for $P \ge 80 \%$.

Our work generalizes previous studies \cite{so2011,skardal2014,ciszak2020, ciszak2021} to the
case of highly-diluted random networks and demonstrates a remarkable
robustness of macroscopic excitability induced by adaptive feedback.
These results establish a bridge between the microscopic and mesoscopic
dynamics, showing how sub-groups of rotators can coherently combine to
produce a macroscopic burst similarly to that of a single excitable
cell. We expect that our work will inspire new research in the study of emergent
phenomena in networks of interacting elements at the mesoscopic scale,
or even in more complex systems such as networks composed by different
kinds of functional units or networks of networks.

\acknowledgments{AT received financial support by the Labex MME-DII (Grant No ANR-11-LBX-0023-01) (together with MP) and by the ANR Project ERMUNDY (Grant No ANR-18-CE37-0014), all part of the French programme ``Investissements d'Avenir''. Part of this work has been developed during the visit of SO during 2021 to the Maison internationale de La Recherche, Neuville-sur-Oise, France supported by CY Advanced Studies, CY Cergy Paris Universit\'e, France.

\vspace{0.5 truecm}

\noindent
{\bf AUTHOR DECLARATIONS}

\vspace{0.5 truecm}
\noindent
{\bf  Conflict of Interest}

The authors have no conflicts to disclose.

\vspace{0.5 truecm}
\noindent
{\bf DATA AVAILABILITY}

The data that support the findings of this study are available
from the corresponding author upon reasonable request


%
 
\end{document}